\documentclass[12pt,oneside,english,12pt]{amsart}
\usepackage[T1]{fontenc}
\usepackage[latin9]{inputenc}
\usepackage{verbatim}
\usepackage{amstext}
\usepackage{amsthm}
\usepackage{amssymb}
\usepackage{graphicx}
\usepackage{esint}

\makeatletter
\numberwithin{equation}{section}
\numberwithin{figure}{section}

\usepackage{amsfonts}
\usepackage[font=small,labelfont=bf]{caption}

\usepackage{babel}

\usepackage{babel}

\makeatother

\usepackage{babel}
\begin{document}
\title{Wilson-Fisher fixed points for any dimension}
\author{R. Trinchero}
\begin{abstract}
The critical behavior of a non-local scalar field theory is studied.
This theory has a non-local quartic interaction term which involves
a real power $-\beta$ of the Laplacian. The parameter $\beta$ can
be tuned so as to make that interaction marginal for any dimension.
The lowest order Feynman diagrams corresponding to coupling constant
renormalization, mass renormalization, and field renormalization are
computed. In all cases a non-trivial IR fixed point is obtained. Remarkably,
for dimensions different from $4$, field renormalization is required
at the one-loop level. For $d=4$, the theory reduces to the usual
local $\phi^{4}$ field theory and field renormalization is required
starting at the the two-loop level. The critical exponents $\nu$
and $\eta$ are computed for dimensions $2,3,4$ and $5$. For dimensions
greater than four, the critical exponent $\eta$ turns out to be negative
for $\epsilon>0$, which indicates a violation of the unitarity bounds.
\end{abstract}

\date{01/04/2019}
\maketitle

\section{Introduction}

The computation of critical exponents for the $3$-dimensional Ising
model using the $\epsilon$-expansion provides a concrete example
of the relevance of the renormalization group ideas\cite{Wilson:1971dc}\cite{Wilson:1973jj}.
This is done by considering a self -interacting $\phi^{4}$ theory
in $d=4-\epsilon$ dimensions, where $\epsilon$ is allowed to take
real values. This procedure led to a qualitative understanding of
the $3$-dimensional Ising model physics, and to predictions for critical
exponents in reasonable agreement with the exact values.

The renormalization group consists in the study of the evolution of
a system under scale transformations. This system involves all possible
interactions of any range for all kinds of dynamical variables. Different
physical systems correspond to the study of particular fixed points
in this huge space of couplings. This paper studies a particular example
of system described near the corresponding fixed point by a non-local
field theory. Non-local field theories appear in various aspects of
physics. These include proposals for dealing with quantum gravity\cite{Barvinsky:2014lja},
field theories based on non-commutative geometry\cite{Blaschke:2016gxl},
and in critical phenomena. The use of non-local field theories in
the description of critical phenomena is not new\cite{Fateev:1985mm},\cite{Reddy2014Auth},\cite{Paulos:2015jfa},
\cite{Egolf2017TheMF}. Such models appear in statistical systems
with long range interactions. In a previous paper\cite{2018PhRvD..98e6023T},
the critical behavior of a $\phi^{4}$ theory with a non-local kinetic
term was studied. That theory has asymptotic freedom in the UV. In
this paper, the critical behavior of a field theory which includes
a non-local interaction is studied. This interaction consists in a
$\phi^{4}$ term which involves a power $-\beta$ of the Laplacian.
The parameter $\beta$ allows to tune the dimensions of the corresponding
coupling constant. This is done in such a way that for any space dimension
$n$ the coupling is a-dimensional. Thus reproducing for any spatial
dimension $n$ what happens for the local $\phi^{4}$ theory in $n=4$
dimensions. The theory is formulated using an auxiliary field $\rho$
, which renders the interaction local, at the cost of having a non-local
kinetic term. The renormalization and critical properties of this
theory are studied at the one-loop level. The features and results
of this work are summarized as follows, 
\begin{itemize}
\item The theory to be considered is a scalar field theory involving a non-local
$\phi^{4}$ interaction term, which includes a power $-\beta$ of
the Laplacian. 
\item The theory is dimensionally regularized, and the contribution of the
one-loop Feynman diagrams which present poles when $\epsilon\to0$
is computed.
\item These one-loop computations show that mass and coupling constant renormalization
are required at this level in order to render the theory finite. In
addition, and contrasting with local $\phi^{4}$ theory, field renormalization
is required at this one-loop level for any dimension different from
$4$.
\item The previous calculation allows to compute the fixed point value for
the coupling constant, and the critical exponents $\nu$ and $\eta$.
In all cases a non-trivial IR fixed point is obtained for $\epsilon>0$.
\item For dimensions $d=n-\epsilon$ with $n$ less than $4$, non-trivial
fixed points are obtained describing consistent quantum theories.
\item For dimensions $d=n-\epsilon$ with $n$ greater than $4$, the critical
exponent $\eta$ turns out to be negative for $\epsilon>0$, which
indicates a violation of the unitarity bounds.
\end{itemize}
The paper is organized as follows. Section 2. presents the model and
explains how to tune $\beta$ in order to get an a-dimensional coupling.
Section 3 contains the computations of Feynman diagrams. Subsection
3.1 presents the one-loop corrections to the two-point function, which
involve the contribution of two diagrams. Subsection 3.2 deals with
the one-loop corrections to the four-point function, which involve
the contribution of three diagrams. Subsection 3.3 computes the critical
exponents $\nu$ and $\eta$ for these theories. Finally, section
4 presents some concluding remarks and additional research motivated
by this work.

\section{The action}

The Euclidean action for the field $\phi$ to be considered is given
by, 
\[
\tilde{S}=\tilde{S}_{0}+\tilde{S}_{I},\:\:\:\tilde{S}_{0}=\int d^{n}x\;\frac{1}{2}\phi(-\triangle+m_{0}^{2})\phi,\;\;\tilde{S}_{I}=\frac{\lambda_{0}}{4!}\int d^{d}x\ensuremath{\phi}^{2}\text{(-\ensuremath{\triangle}+\ensuremath{M^{2})^{-\beta}\ensuremath{\phi}^{2}}}\;
\]
In what follows, bare mass and coupling will be indicated by $m_{0}$
and $\lambda_{0}$ , the corresponding renormalized quantities will
be $m$ and $\lambda$. Using an auxiliary field $\rho$, the action
$\tilde{S}$ can be replaced by the following equivalent action,
\[
S=S_{0}+S_{I},\;\;\;S_{0}=\int d^{n}x\;\left[\frac{1}{2}\phi(-\triangle+m_{0}^{2})\phi\:-\frac{1}{2}\rho(-\triangle+M^{2})^{\beta}\rho\right],\;\;S_{I}=-\int d^{n}x\;\sqrt{\frac{\lambda_{0}}{12}}\rho\ensuremath{\phi}^{2}
\]
functionally integrating over $\rho$ leads back to the action $\tilde{S}$
, i.e
\[
\int\mathcal{D}\rho\;e^{-S}\propto e^{-\tilde{S}}
\]
The free energy is given by,
\[
F=-\log Z\;\;,Z=\int\mathcal{D}\phi\,\mathcal{D}\rho\,e^{-S}
\]
The Fourier transform of the free $\rho$ and $\phi$ two-point functions
are therefore given by, 
\begin{eqnarray*}
<\rho\rho>(p)=-\frac{1}{(p^{2}+M^{2})^{\beta}} & \,, & <\phi\phi>(p)=\frac{1}{(p^{2}+m_{0}^{2})}
\end{eqnarray*}
the mass $M$ has been included in order to regulate infrared divergencies.

It is noted that the dimension of the field $\phi$ and coupling $[\lambda_{0}]$
are,
\[
[\phi]=\frac{d-2}{2}\;\;,\;\;[\lambda_{0}]=d-4[\phi]+2\beta=4-d+2\beta
\]
for $d=n-\epsilon$ choosing,
\begin{equation}
\beta=\frac{n}{2}-2\label{eq:beta}
\end{equation}
makes the coupling dimension equal to $\epsilon,$ i.e. $[\lambda_{0}]=\epsilon$.
This choice will be adopted from now on. This is exactly what happens
for the coupling of the local $\phi^{4}$ theory in $d=4-\epsilon$.
Thus the non-local coupling parametrized by $\beta$ in $\tilde{S}_{I}$
can be considered as a device to reproduce the same situation in other
dimensions. It is also noted that for $n>4$ then $\beta>0$, which
for $M\to0$ produces infrared divergencies in these cases. This motivates
the choice $M=M_{n}$ , where,
\begin{equation}
M_{n}=\left\{ \begin{array}{cc}
0 & n\leq4\\
M & n>4
\end{array}\right.\label{eq:Mn}
\end{equation}

\section{Effective Couplings}

\subsection{One-loop correction to the two-point function.}

The one-loop corrections to the two-point function are given by the
following diagrams, 
\begin{center}
\includegraphics[scale=0.3]{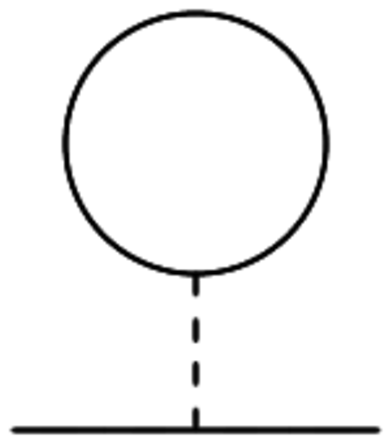}\includegraphics[scale=0.3]{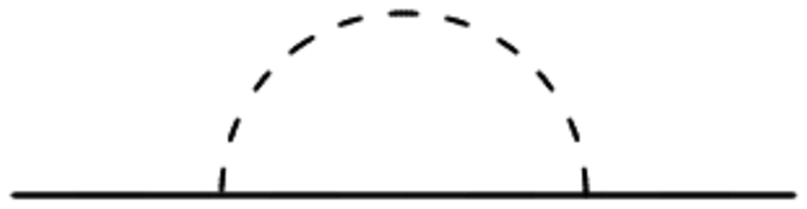}
\label{-figure-2-point}\captionof{figure}{Lowest order diagrams
contributing to renormalization of the $\phi$ two-point function.} 
\par\end{center}

\noindent It is noted that the internal $\rho$ line in the first
diagram carries zero momentum. Thus, taking into account the relation
(\ref{eq:beta}), this line gives a factor,
\[
\frac{-1}{M_{n}^{2\beta}}=\frac{-1}{M_{n}^{n-4}}
\]
which recalling (\ref{eq:Mn}) shows that this line vanishes for $n<4$,
gives $-1$ for $n=4$ and is non-vanishing for $n>4$. Indeed the
motivation for introducing the mass $M_{n}$ for the $\rho$ field,
was to regulate the infrared divergence appearing from this propagator
at zero momentum for $n>4,$ i.e. $\beta>0$. The contribution of
the first diagram without external legs is, 
\begin{align}
a_{T}(d,\beta) & =\mu_{T}\frac{-\lambda_{0}}{2!\,12}M_{n}^{4-n}I_{T}(d)=\frac{-\lambda_{0}}{6}M_{n}^{4-n}I_{T}(d)\label{eq:ait}\\
I_{T}(d) & =\intop\frac{d^{d}q}{(2\pi)^{d}}\frac{1}{(q^{2}+m_{0}^{2})}\nonumber 
\end{align}
the quantity $\mu_{T}=4$ is the multiplicity of this diagram. Introducing
the Feynman parametrization and integrating over the momenta $q$,
leads to, %
\begin{align*}
I_{T}(d) & =\frac{\Gamma(1-\frac{d}{2})}{(4\pi)^{\frac{d}{2}}}m_{0}^{2(\frac{d}{2}-1)}
\end{align*}
replacing in (\ref{eq:ait}) gives, 
\begin{align*}
a_{T}(d,\beta) & =-\frac{\lambda_{0}}{6}M_{n}^{4-n}\frac{\Gamma(1-\frac{d}{2})}{(4\pi)^{\frac{d}{2}}}m_{0}^{2(\frac{d}{2}-1)}\\
 & =-\frac{g_{0}}{6}\mu^{2(\beta+2)-d}M_{n}^{4-n}\frac{\Gamma(1-\frac{d}{2})}{(4\pi)^{\frac{d}{2}}}m_{0}^{2(\frac{d}{2}-1)}
\end{align*}
where in the second equality the result has been expressed in terms
of the a-dimensional coupling $g_{0}=\mu^{d-2(\beta+2)}\lambda_{0}$
. For $n$ odd, this contribution presents no divergence for $\epsilon\to0$.
For $d=n-\epsilon$ with $n$ a positive even integer, the gamma function
can be expanded as follows,
\[
\Gamma(1-\frac{d}{2})=\Gamma(1-\frac{d}{2})=\frac{(-1)^{\frac{n}{2}-1}}{(\frac{n}{2}-1)!}\frac{2}{\epsilon}+R
\]
where $R$ denotes terms that present no singularities when $\epsilon\to0$.
This leads to,
\[
a_{T}(n-\epsilon,\beta)=-\frac{g_{0}}{6}\mu^{2(\beta+2)-n+\epsilon}M_{n}^{4-n}\frac{m_{0}^{2(\frac{n}{2}-1-\frac{\epsilon}{2})}}{(4\pi)^{\frac{d}{2}}}\frac{(-1)^{\frac{n}{2}-1}}{(\frac{n}{2}-1)!}\frac{2}{\epsilon}+R'
\]

\noindent For $n=2$ and $3$ , because of the remarks at the beginning
of this subsection, this diagram gives no contribution, which is also
the case for any odd $n$. Thus for $n<6$ the singular part of $a_{T}$
is given by,
\[
a_{T}^{SP}(n-\epsilon,\frac{n}{2}-2)=\left\{ \begin{array}{cc}
0 & n=2\\
0 & n=3\\
\frac{g_{0}}{3}\frac{m_{0}^{2}}{(4\pi)^{2}}\frac{1}{\epsilon} & n=4\\
0 & n=5
\end{array}\right.
\]

\noindent The integral corresponding to the second diagram is,
\begin{align*}
\Pi(p,m_{0},d,\beta) & =-\mu_{\text{\ensuremath{\Pi}}}\frac{\lambda_{0}}{2!\,12}I_{F}(p,m_{0},d,\beta)=-\frac{\lambda_{0}}{3}I_{F}(p,m_{0},d,\beta)\\
I_{F}(p,m_{0},d,\beta) & =\int\,\frac{d^{d}q}{(2\pi)^{d}}\frac{1}{\left(q^{2}+m_{0}^{2}\right)\left((p-q)^{2}+M_{n}^{2}\right)^{\beta}}
\end{align*}
the quantity $\mu_{\Pi}=8$ is the multiplicity of this diagram. The
degree of divergence of this integral is,
\[
\omega(I_{F})=d-2-2\beta
\]
for the cases considered in this work ($d=n-\epsilon$, $\beta=\frac{n}{2}-2$)
this gives $\omega(I_{F})=2-\epsilon$ . This implies that there is
a divergent term proportional to $p^{2}$, i.e. that this diagram
requires a field renormalization. Below it is shown that the coefficient
of $p^{2}$ vanishes for $d=4$, this should be so because for $n=4$
, $\beta=0$ and the theory coincides with the usual local $\phi^{4}$
field theory, which requires no field renormalization at the one-loop
level. Introducing the Feynman parametrization and integrating over
the momenta $q$, leads to, 
\begin{align*}
I_{F}(p,m_{0},d,\beta) & =\frac{\mu^{2(-\beta+\frac{d}{2}-1)}}{(4\pi)^{\frac{d}{2}}}\frac{\Gamma\left(-\frac{d}{2}+\beta+1\right)}{\Gamma(\beta)}\int_{0}^{1}dx\,(1-x)^{\beta-1}\\
 & \times\left(\frac{M_{n}^{2}}{\mu^{2}}(1-x)+\frac{m_{0}^{2}}{\mu^{2}}x+\frac{p^{2}}{\mu^{2}}(1-x)x\right)^{-\beta+\frac{d}{2}-1}
\end{align*}
The power of the parenthesis in the integrand is $-\beta+\frac{d}{2}-1=1-\frac{\epsilon}{2}$,
which therefore presents no singularities in the integration region.
Thus $\epsilon=0$ is taken in this power. The integral can be evaluated
leading to,%
\[
I_{F}(p,m_{0},d,\beta)=\frac{\mu^{-\epsilon}}{(4\pi)^{\frac{d}{2}}}\frac{\Gamma\left(-\frac{d}{2}+\beta+1\right)}{\Gamma(\beta)}\frac{\left(\beta\left((\beta+2)M_{n}^{2}+p^{2}\right)+(\beta+2)m_{0}^{2}\right)}{\beta(\beta+1)(\beta+2)}
\]
which leads to the following expression for its singular part,
\begin{equation}
I_{F}^{SP}(p,m_{0},d,\beta)=-\frac{2^{2-n}\pi^{-n/2}\left(M^{2}n^{2}-4M^{2}n+2m_{0}^{2}n+2np^{2}-8p^{2}\right)}{(n-4)(n-2)n\Gamma\left(\frac{n}{2}-2\right)\;\epsilon}\label{eq:self-phi-gral}
\end{equation}
This leads to,
\begin{align*}
\Pi^{SP}(p,m_{0},d,\beta)=-\frac{g_{0}}{3}I_{F}^{SP}(p,m_{0},d,\beta) & =\left\{ \begin{array}{cc}
\frac{g_{0}}{4\pi}\frac{1}{3}\left(\frac{2}{\epsilon}\right)(m_{0}^{2}-p^{2}) & n=2\\
\frac{g_{0}}{8\pi^{2}}\frac{1}{9}\left(\frac{2}{\epsilon}\right)(6m_{0}^{2}-2p^{2}) & n=3\\
\frac{g_{0}}{(4\pi)^{2}}\frac{1}{3}\left(\frac{2}{\epsilon}\right)m_{0}^{2} & n=4\\
\frac{g_{0}}{\pi^{3}}\frac{(5M^{2}+2(5m_{0}^{2}+p^{2}))}{360}\frac{1}{\epsilon} & n=5
\end{array}\right.
\end{align*}
this result is quite remarkable because it shows that for $n\neq4\Rightarrow\beta\neq0$
field renormalization is required at the one-loop level to absorb
the divergences. It is recalled that for the $n=4$ local $\phi^{4}$
theory, field renormalization is first required at two loops. The
diagram contributing to field renormalization is the second one in
fig. \ref{-figure-2-point}, and it requires field renormalization
because of the non-zero momentum carried by the auxiliary field $\rho$
line. Alternatively this can be seen from the fact that the non-local
interaction introduces momentum at each non-local vertex. It is remarked
that the general result (\ref{eq:self-phi-gral}) shows that starting
from $n=5$ the sign of the term proportional to $p^{2}$ changes,
this sign change will affect the sign of the critical exponent $\eta$,
leading to a violation of the unitarity bounds for $n\geq5$.

\noindent The effective two-point function to one-loop order is,\footnote{The free energy is,
\[
F=-\log Z\;\;,Z=\int\mathcal{D}\phi\,\mathcal{D}\rho\,e^{-S}
\]
therefore the order $0$ free energy goes as $S_{0}$, and the corrections
carry a minus sign in front.}, 
\begin{align}
\Gamma_{2}(p) & =p^{2}+m_{0}^{2}-\left(a_{T}^{SP}+\Pi^{SP}\right)\nonumber \\
 & =p^{2}Z_{\phi}^{-1}+m_{0}^{2}Z_{m^{2}}^{-1}=Z_{\phi}^{-1}\left(p^{2}+m_{0}^{2}\frac{Z_{\phi}}{Z_{m^{2}}}\right)\label{eq:field-mass-renor}
\end{align}
where,
\begin{align*}
Z_{\phi} & =1-\frac{g_{0}}{(4\pi)^{\frac{n}{2}}}\frac{1}{\epsilon}c_{\phi}(n)\;\;,\;\;Z_{m^{2}}=1+\frac{g_{0}}{(4\pi)^{\frac{n}{2}}}\frac{1}{\epsilon}c_{m^{2}}(n)
\end{align*}
with,
\[
c_{\phi}(n)=\left\{ \begin{array}{cc}
\frac{2}{3} & n=2\\
\frac{4}{9}\frac{1}{\sqrt{\text{\ensuremath{\pi}}}} & n=3\\
0 & n=4\\
-\frac{32}{180}\frac{1}{\sqrt{\text{\ensuremath{\pi}}}} & n=5
\end{array}\right.\;,\;c_{m^{2}}(n)=\left\{ \begin{array}{cc}
\frac{2}{3} & n=2\\
\frac{4}{3\sqrt{\pi}} & n=3\\
1 & n=4\\
\frac{32}{36\sqrt{\pi}}(1+\frac{M_{n}^{2}}{2m_{0}^{2}}) & n=5
\end{array}\right.
\]
 The effective mass is therefore given by,
\[
m^{2}=m_{0}^{2}\frac{Z_{\phi}}{Z_{m^{2}}}
\]
\footnote{It is noted that for $d=4-\epsilon$ the expression for the effective
mass is,
\[
m^{2}=m_{0}^{2}-\frac{g_{0}m_{0}^{2}}{(4\pi)^{2}}\frac{1}{\epsilon}\left(\frac{1}{3}+\frac{2}{3}\right)=m_{0}^{2}-\frac{g_{0}m_{0}^{2}}{(4\pi)^{2}}\frac{1}{\epsilon}
\]

which is the well-known result for the local $\phi^{4}$ theory.}

\noindent The gamma function is given by,
\begin{align*}
\gamma(g) & =\mu\left.\frac{\partial}{\partial\mu}\log Z_{\phi}^{\frac{1}{2}}\right|_{\lambda\,fixed}=\frac{1}{2}\frac{\partial}{\partial\log\mu}\log\left(1-\frac{\lambda_{0}\mu^{-\epsilon}}{(4\pi)^{\frac{n}{2}}}\frac{1}{\epsilon}c_{\phi}(n)\right)\\
 & =\frac{1}{2\,Z_{\phi}}\left(\frac{g_{0}}{(4\pi)^{\frac{n}{2}}}c_{\phi}(n)\right)=\frac{1}{2\,}\frac{g_{0}}{(4\pi)^{\frac{n}{2}}}c_{\phi}(n)
\end{align*}
thus,
\[
\gamma(g)=\left\{ \begin{array}{cc}
\frac{1}{3\,}\frac{g_{0}}{4\pi} & n=2\\
\frac{2}{9}\frac{1}{\sqrt{\text{\ensuremath{\pi}}}}\frac{g_{0}}{(4\pi)^{\frac{3}{2}}} & n=3\\
0 & n=4\\
-\frac{16}{180\sqrt{\text{\ensuremath{\pi}}}\,}\frac{g_{0}}{(4\pi)^{\frac{5}{2}}} & n=5
\end{array}\right.
\]

\subsection{One-loop correction to the four-point function}

\subsubsection{Vertex correction\label{subsec:-vertex-correction}}

The diagram to be considered is given by the figure below,

\hspace{6cm}\includegraphics[scale=0.25]{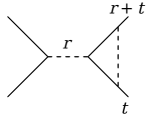} \captionof{figure}{
$\rho\phi^{2}$ vertex correction contributing to the 4 point $\phi$
function.} 

\noindent Removing external legs, the integral to be computed is,
\begin{align*}
I_{V}(r,t,m_{0},M_{n},d,\beta) & =\frac{\mu_{V}}{4!}\left(\frac{\lambda_{0}}{12}\right)^{2}(r^{2}+M_{n}^{2})^{-\beta}A_{V}(r,t,m_{0},M_{n},d,\beta)\\
 & =\frac{2}{3}\lambda_{0}^{2}(r^{2}+M_{n}^{2})^{-\beta}A_{V}(r,t,m_{0},M_{n},d,\beta)\\
A_{V}(r,t,m_{0},M_{n},d,\beta) & =\int\,\frac{d^{d}q}{(2\pi)^{d}}\frac{1}{\left(q^{2}+m_{0}^{2}\right)\left((q-r)^{2}+m_{0}^{2}\right)\left((q-r-t)^{2}+M_{n}^{2}\right)^{\beta}}
\end{align*}
the quantity $\mu_{V}=96\,4!$ is the multiplicity of this diagram.
The overall factor $\frac{2}{3}$ in front of this contribution comes
from,
\[
\frac{\mu_{V}}{4!12^{2}}=\frac{2}{3}
\]
 The degree of divergence of this integral is,
\[
\omega(A_{V})=d-4-2\beta
\]
for the cases considered in this work ($d=n-\epsilon$, $\beta=\frac{n}{2}-2$)
this gives $\omega(A_{V})=d-4-2\beta=-\epsilon$ . Therefore, taking
derivatives with respect to $M_{n}^{2}$ makes the integral convergent,
thus making a series expansion in powers of $M_{n}^{2}$, only the
first term can produce a divergence when $\epsilon\to0,$ i.e. $M_{n}=0$
is taken in the calculation below. The same reasoning applies for
the dependence of $A_{V}(r,t,m_{0},M_{n},d,\beta)$ on the external
momenta $r$ and $t$, thus only zero external momenta is relevant
for the singular part of this contribution. This leads to,
\[
A_{V}(0,0,m_{0},0,d,\beta)=\int\,\frac{d^{d}q}{(2\pi)^{d}}\frac{1}{\left(q^{2}+m_{0}^{2}\right)\left(q^{2}+m_{0}^{2}\right)\left(q^{2}\right)^{\beta}}
\]
Introducing the Feynman parametrization and integrating over the moment
$q$, leads to, 
\begin{align*}
A_{V}(0,0,m_{0},0,d,\beta) & =\frac{\Gamma\left(-\frac{d}{2}+\beta+2\right)}{(4\pi)^{\frac{d}{2}}\Gamma(\beta)}\int_{0}^{1}dy\,y^{\beta-1}(1-y)\left(m_{0}^{2}(1-y)\right)^{\frac{1}{2}(d-2(\beta+2))}\\
 & =\frac{\Gamma\left(-\frac{d}{2}+\beta+2\right)}{(4\pi)^{\frac{d}{2}}\Gamma(\beta)}\int_{0}^{1}dy\,y^{\beta-1}(1-y)^{1+\frac{1}{2}(d-2(\beta+2))}m_{0}^{d-2(\beta+2)}\\
 & =\frac{\Gamma\left(\frac{d}{2}-\beta\right)\Gamma\left(-\frac{d}{2}+\beta+2\right)}{(4\pi)^{\frac{d}{2}}\Gamma\left(\frac{d}{2}\right)}\mu^{d-2(\beta+2)}\left(\frac{m_{0}}{\mu}\right)^{d-2(\beta+2)}\\
 & \overset{_{d=n-\epsilon,\beta=\frac{n}{2}-2}}{=}\frac{\Gamma\left(2-\frac{\epsilon}{2}\right)\Gamma\left(\frac{\epsilon}{2}\right)}{(4\pi)^{\frac{d}{2}}\Gamma\left(\frac{n-\epsilon}{2}\right)}\mu^{-\epsilon}\left(\frac{m_{0}}{\mu}\right)^{-\epsilon}
\end{align*}
whose pole term in $\epsilon$ is,%
\[
A_{V}(0,0,m_{0},0,d,\beta)=\frac{\mu^{-\epsilon}}{(4\pi)^{\frac{d}{2}}}\frac{2}{\Gamma\left(\frac{n}{2}\right)}\frac{1}{\epsilon}
\]
leading to,
\[
I_{V}(r,t,m_{0},M,d,\beta)=\frac{2}{3}\lambda_{0}^{2}(r^{2}+M^{2})^{-\beta}\frac{\mu^{-\epsilon}}{(4\pi)^{\frac{d}{2}}}\frac{2}{\Gamma\left(\frac{n}{2}\right)}\frac{1}{\epsilon}
\]

\subsubsection{$\rho$-self energy correction}

The diagram to be considered is the following,
\begin{center}
\includegraphics[scale=0.2]{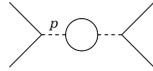} \captionof{figure}{
$\text{\ensuremath{\rho}}$ self-energy correction contributing to
the four-point $\phi$ function.} 
\par\end{center}

\noindent The integral to be computed is,
\begin{align*}
I_{\rho}(p,m_{0},d) & =\frac{\mu_{\rho}}{4!}\left(\frac{\lambda_{0}}{12}\right)^{2}\frac{1}{(p^{2}+M^{2})^{2\beta}}\,A(p,m_{0},d)=\frac{2}{3}\lambda_{0}^{2}(p^{2}+M^{2})^{-2\beta}\,A(p,m_{0},d)\\
A(p,m_{0},d) & =\int\,\frac{d^{d}q}{(2\pi)^{d}}\frac{1}{\left(q^{2}+m_{0}^{2}\right)\left((p-q)^{2}+m_{0}^{2}\right)}
\end{align*}
for this diagram $\mu_{\rho}=96\,4!$ leading to the same overall
factor $\frac{2}{3}$ , as in the previous diagram. Introducing the
Feynman parametrization and integrating over $q$, leads to, 
\begin{align*}
A(p,m_{0},d) & =\frac{\Gamma\left(2-\frac{d}{2}\right)}{(4\pi)^{\frac{n}{2}}}\int_{0}^{1}dx\frac{1}{\left(m_{0}^{2}+p^{2}x(1-x)\right)^{2-\frac{d}{2}}}
\end{align*}
the integral over $x$ gives a finite result for any $n$, since the
integrand is well-behaved in the entire integration region. Therefore
$\Gamma\left(2-\frac{d}{2}\right)$ is the one that can produce a
pole for $\epsilon\to0$, this does not happen for $d=n-\epsilon$
with $n$ odd neither for $n=2$. Thus, only even values of $n$ with
$n\geq4$ produce poles. For $n=4$ the original integral is logarithmically
divergent, and therefore in getting the pole term, only the value
of the integral for $p^{2}=0$ should be computed, this gives,%
\[
A(p,m_{0},4-\epsilon)=\frac{\Gamma\left(\frac{\epsilon}{2}\right)}{(4\pi)^{\frac{n}{2}}}(m_{0}^{2})^{-\frac{\epsilon}{2}}+C''=\frac{(m_{0}^{2})^{-\frac{\epsilon}{2}}}{(4\pi)^{\frac{n}{2}}}\frac{2}{\epsilon}+C'''
\]
For $n\geq6$ even, the integral $A(p,m_{0},d)$ has degree of divergence
greater than zero, given by $\omega(A)=n-\epsilon-4$ . This means
that the derivatives of $A(p,m_{0},d)$ with respect to $p^{2}$ have
divergent coefficients when $\epsilon\to0$ up to order $\omega(A)/2$.
The number of these divergent coefficients matches the number of parameters
needed to specify the $\rho$ free field action. Indeed, for $n=6$
then $\beta=1$ which means that the coefficient of $\rho^{2}$ and
the coefficient of $\rho\triangle\rho$ get renormalized in this case.
The cases with $n\geq6$ even will not be further studied in this
work.

\noindent Thus for $n<6$ the following result for the pole term of
$I_{\rho}(p,m_{0},d)$ is obtained,
\[
I_{\rho}(p,m_{0},d)=\delta_{n,4}\frac{2}{3}\lambda_{0}^{2}\frac{\mu^{-\epsilon}}{(4\pi)^{\frac{n}{2}}}\frac{2}{\epsilon}
\]

\subsubsection{Box}

The diagram to be considered is the following, 
\begin{center}
\includegraphics[scale=0.2]{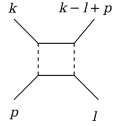} \captionof{figure}{ Box diagram
contributing to the four-point $\phi$ function.}
\par\end{center}

\noindent The integral to be computed is,
\begin{align*}
I_{B}(p,m_{0},d) & =\frac{\mu_{B}}{4!}\left(\frac{\lambda_{0}}{12}\right)^{2}A_{B}(p,k,l,m_{0},d)\\
 & =\frac{1}{6}\lambda_{0}^{2}A_{B}(p,k,l,m_{0},d)
\end{align*}
where $\mu_{B}=24\,4!$ denotes the multiplicity of this contribution
leading to the overall factor $\frac{1}{6}.$The quantity $A_{B}(p,k,l,m_{0},d)$
is,
\begin{align*}
A_{B}(p,k,l,m_{0},d) & =\int\,\frac{d^{d}q}{(2\pi)^{d}}\frac{1}{\left((k-q)^{2}+m_{0}^{2}\right)\left(q^{2}+M^{2}\right)^{\beta}}\\
 & \times\frac{1}{\left((q+p)^{2}+m_{0}^{2}\right)\left((q+p-l)^{2}+M^{2}\right)^{\beta}}
\end{align*}
 The degree of divergence of this integral is,
\[
\omega_{B}=d-4-4\beta
\]
which for the cases considered in this work $d=n-\epsilon,\beta=\frac{n}{2}-2$
is,
\[
\omega_{B}=4-n-\epsilon
\]
This shows that for $n>4$ there is no divergence for $\epsilon\to0$.
Introducing the Feynman parametrization , leads to, 
\begin{align}
A_{B}(p,k,l,m_{0},d) & =\int_{0}^{1}dx\,dy\,dz\frac{(1-x)^{\beta-1}y^{\beta-1}((1-z)z)^{\beta}\Gamma(2+2\beta)}{(4\pi)^{\frac{d}{2}}\Gamma(\beta)^{2}}\nonumber \\
 & \times\mu^{4-n-\epsilon}P(k,p,l,x,y,z)\label{eq:ab}
\end{align}

\noindent where,
\begin{align*}
P(k,p,l,x,y,z) & =\int\,\frac{d^{d}q}{(2\pi)^{d}}\left((1-z)\left((1-y)\left((k-q)^{2}+m_{0}^{2}\right)+\right.\right.\\
 & +\left.\left((1-z)\left(y\left((q+p-l)^{2}+M^{2}\right)\right)\right.\right)+\\
 & +\left.z\left(x\left((q+p)^{2}+m_{0}^{2}\right)-(x-1)\left(q+M^{2}\right)^{2}\right)\right)^{2-n}
\end{align*}
is positive definite in the $x,y,z$ integration region. The integral
in the last expression can be done leading to,
\begin{align*}
P(k,p,l,x,y,z)= & \frac{\Gamma\left(-\frac{d}{2}+2\beta+2\right)}{\Gamma(2+2\beta)}\left(k^{2}(y-1)(z-1)-\left(k(y(1-z)+z-1)\right.\right.\\
 & \left.+M^{2}(1-x)z+p(xz+y(1-z))+qyz-qy\right)^{2}+M^{4}(1-x)z\\
 & -M^{2}yz+M^{2}y+m_{0}^{2}((x-1)z+y(z-1)+1)+p^{2}xz\\
 & \left.+p^{2}y(1-z)+2pqyz-2pqy+q^{2}y(1-z)\right)^{\frac{1}{2}(d-4(\beta+1))}
\end{align*}
it is noted that for $n\leq4$ the power appearing in $P(k,p,l,x,y,z)$
is positive or goes to zero for $\epsilon\to0$. In addition, $P(k,p,l,x,y,z)$
is well defined and positive in the $x,y,z$ integration region. Thus,
the integral in (\ref{eq:ab}) is bounded from above by the integral
of $P(k,p,l,x,y,z)$ multiplied by the following positive integral,
\begin{align*}
\mu^{4-n-\epsilon}\int_{0}^{1}dx\,dy\,dz\frac{(1-x)^{\beta-1}y^{\beta-1}(-(z-1)z)^{\beta}}{(4\pi)^{\frac{d}{2}}\Gamma(\beta)^{2}} & =\\
=\frac{\mu^{4-n-\epsilon}}{(4\pi)^{\frac{d}{2}}\Gamma(2\beta+2)}\overset{_{d=n-\epsilon,\beta=\frac{n}{2}-2}}{=}\frac{\mu^{4-n-\epsilon}}{(4\pi)^{\frac{d}{2}}\Gamma(n-2)}
\end{align*}
therefore the singularities of $A_{B}(p,k,l,m_{0},d)$ for $\epsilon\to0$
only arise from the factor,
\[
\frac{\Gamma\left(-\frac{d}{2}+2\beta+2\right)}{\Gamma(n-2)}=\frac{\Gamma\left(\frac{n}{2}+\frac{\epsilon}{2}-2\right)}{\Gamma(n-2)}
\]
this produces singularities only for $n=4$. For this value of $n$
and in the limit $\epsilon\to0$, $P(k,p,l,x,y,z)/\Gamma\left(-\frac{d}{2}+2\beta+2\right)$
can be replaced by $1$, leading to,
\begin{align*}
A_{B}(m_{0},n-\epsilon) & =\mu^{4-n-\epsilon}\int_{0}^{1}dx\,dy\,dz\frac{(1-x)^{\beta-1}y^{\beta-1}(-(z-1)z)^{\beta}\Gamma\left(-\frac{d}{2}+2\beta+2\right)}{(4\pi)^{\frac{d}{2}}\Gamma(\beta)^{2}}\\
 & =\mu^{4-n-\epsilon}\frac{\Gamma\left(-\frac{d}{2}+2\beta+2\right)}{(4\pi)^{\frac{d}{2}}\Gamma(2\beta+2)}\overset{_{d=n-\epsilon,\beta=\frac{n}{2}-2}}{=}\mu^{4-n-\epsilon}\frac{\Gamma\left(\frac{n}{2}+\frac{\epsilon}{2}-2\right)}{(4\pi)^{\frac{d}{2}}\Gamma(n-2)}
\end{align*}
which presents a pole term in $\epsilon$ only for $n=4$, given by,%
\[
A_{B}^{PT}(p,k,l,m_{0},4-\epsilon)=\frac{1}{8\pi^{2}\epsilon}
\]
thus,
\[
I_{B}^{PT}(p,m_{0},d)=\frac{1}{6}\lambda_{0}^{2}\frac{\mu^{-\epsilon}}{(4\pi)^{2}}\frac{2}{\epsilon}\delta_{n,4}
\]

\subsubsection{Effective quartic coupling and its $\beta$-function}

\noindent Summing up the contributions from the $\rho\phi^{2}$ vertex
correction, the $\rho$-self energy correction and the box and including
the field renormalization in (\ref{eq:field-mass-renor}), leads to,
\[
\lambda=\frac{\lambda_{0}Z_{\phi}^{2}}{Z_{g}}
\]
where,
\begin{align*}
Z_{\phi} & =1-\frac{g_{0}}{(4\pi)^{\frac{n}{2}}}\frac{1}{\epsilon}c_{\phi}(n)\\
Z_{g} & =1+\frac{g_{0}}{(4\pi)^{\frac{n}{2}}}\frac{1}{\epsilon}c_{g}(n)
\end{align*}

\noindent where,
\[
c_{g}(n)=\frac{4}{3}\frac{1}{\Gamma\left(\frac{n}{2}\right)}+\delta_{n,4}\frac{4}{3}+\delta_{n,4}\frac{1}{3}
\]
the beta function corresponding to the a-dimensional coupling $g,$
is,
\begin{align*}
\beta(g) & =\mu\frac{d}{d\mu}g=\mu\frac{d}{d\mu}\left(\frac{\lambda_{0}Z_{\phi}^{2}}{Z_{g}}\mu^{-\epsilon}\right)\\
 & =-\epsilon g+2gZ_{\phi}^{-1}\mu\frac{d}{d\mu}Z_{\phi}-gZ_{g}^{-1}\mu\frac{d}{d\mu}Z_{g}\\
 & =-\epsilon g+4g\gamma-\frac{c_{g}(n)}{(4\pi)^{\frac{n}{2}}\epsilon}g\beta(g)
\end{align*}
which leads to,
\begin{align}
\beta(g) & =-\epsilon g+4g\gamma+g^{2}\frac{c_{g}(n)}{(4\pi)^{\frac{n}{2}}}\nonumber \\
 & =-\epsilon g+\frac{g^{2}}{(4\pi)^{\frac{n}{2}}}\left(2c_{\phi}(n)+c_{g}(n)\right)\label{eq:betae}
\end{align}
the fixed points defined by $\beta(g^{\star})=0$ are given by,
\[
g_{0}^{\star}=0\;\;,g_{1}^{\star}=\frac{\epsilon(4\pi)^{\frac{n}{2}-\frac{\epsilon}{2}}}{2c_{\phi}(n)+c_{g}(n)}
\]

\noindent For $n=2,3,4,5$, $C(n)>0$, so that for $\epsilon>0$,
$g_{1}^{\star}$ is positive and corresponds to a IR fixed point ,
also for $\epsilon>0$, $g_{0}^{\star}$ corresponds to a UV fixed
point. The figures below shows these $\beta$-functions,

\noindent \hspace{3cm}\includegraphics[scale=0.2]{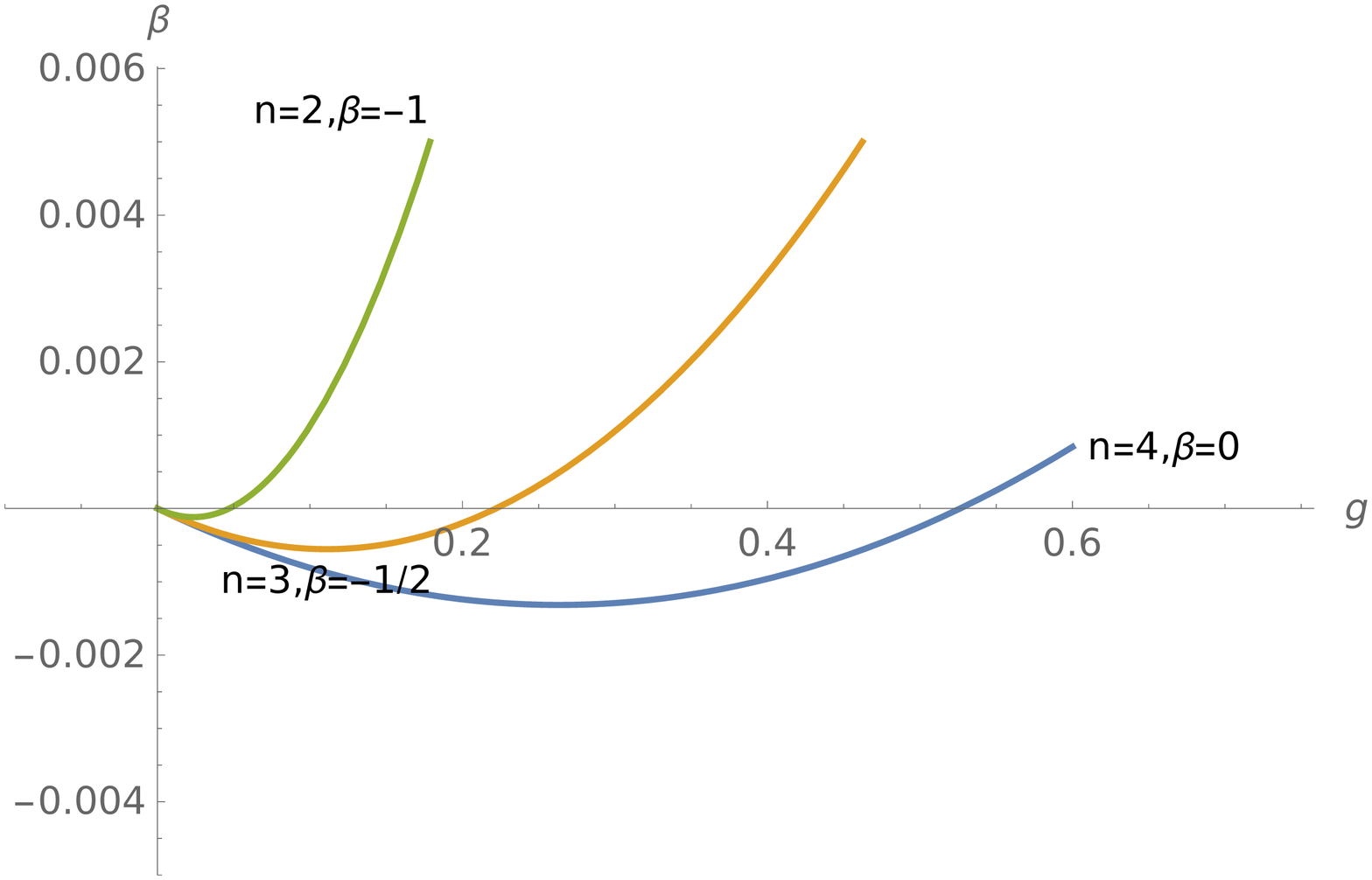}
\includegraphics[scale=0.2]{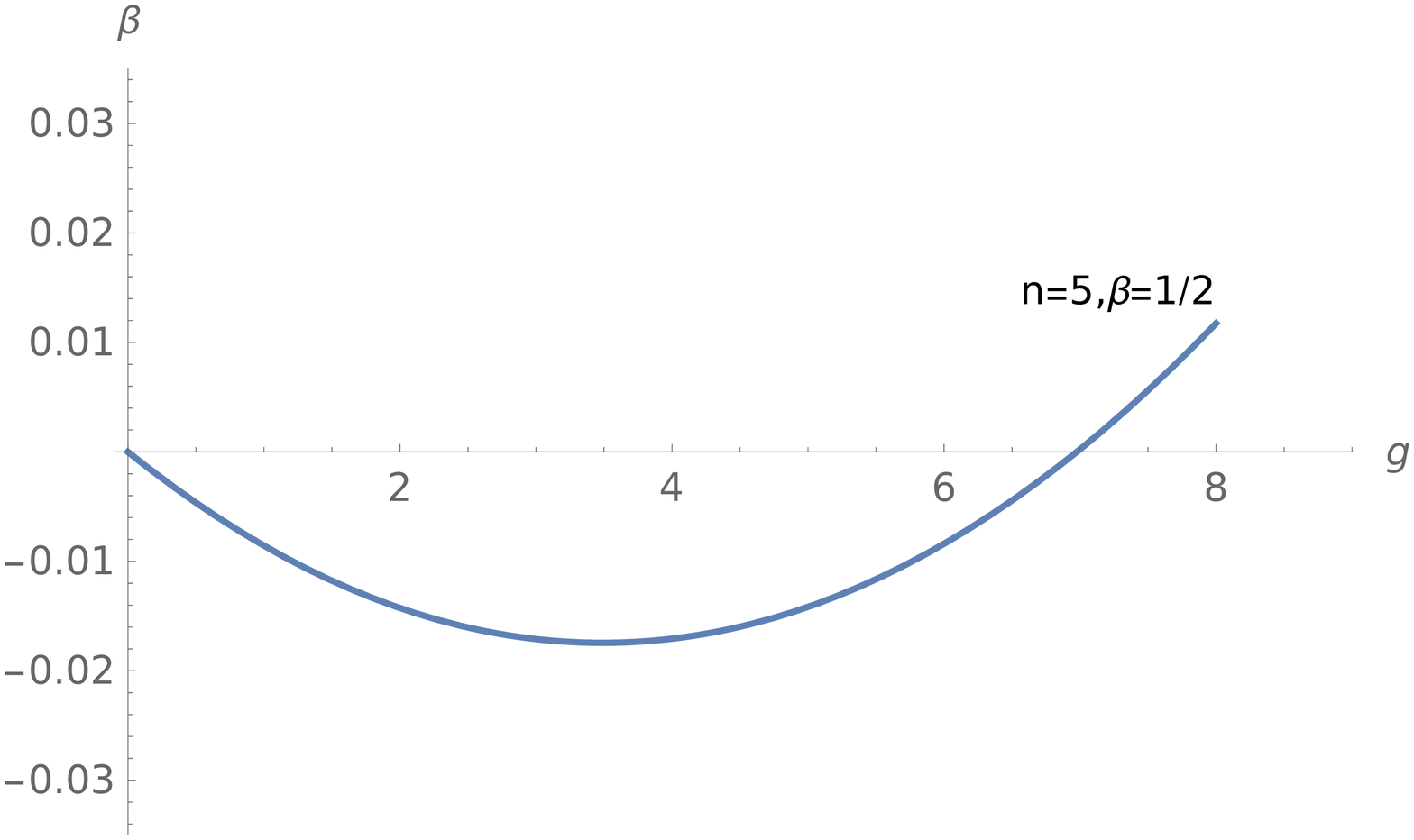}\captionof{figure}{
Beta functions for $\epsilon=0.01$ and dimensions $2,3,4$ and $5$.}

\subsection{The critical exponents $\eta$ and $\nu$}

\noindent The $\gamma_{m}(n)$ function is given by, 
\[
\gamma_{m}(n)=\gamma-\frac{c_{m^{2}}(n)}{(4\pi)^{\frac{n}{2}}}\frac{\beta(g)}{2\epsilon}=\frac{1}{2\,}\frac{g}{(4\pi)^{\frac{n}{2}}}c_{\phi}(n)+\frac{c_{m^{2}}(n)}{(4\pi)^{\frac{n}{2}}}\frac{g}{2}=\frac{1}{2\,}\frac{g}{(4\pi)^{\frac{n}{2}}}\left(c_{\phi}(n)+c_{m^{2}}(n)\right)
\]
where (\ref{eq:betae}) was employed. The critical exponents are related
to the fixed point values $\gamma^{\star}$ and $\gamma_{m}^{\star}$
of the functions $\gamma$ and $\gamma_{m}$. They are given by, 
\[
\nu(n)=\frac{1}{2-2\gamma_{m}^{\star}(n)}\;\;\;,\eta(n)=2\gamma^{\star}(n)
\]
The non-trivial fixed point is given by, 
\[
g_{1}^{\star}(n)=\frac{\epsilon(4\pi)^{\frac{n}{2}}}{2c_{\phi}(n)+c_{g}(n)}
\]
the fixed point values $\gamma_{m}^{\star}(n)$ and $\gamma^{\star}(n)$
are therefore given by, 
\begin{align*}
\gamma_{m}^{\star}(n) & =\frac{1}{2\,}\frac{g_{1}^{\star}(n)}{(4\pi)^{\frac{n}{2}}}\left(c_{\phi}(n)+c_{m^{2}}(n)\right)=\frac{\epsilon}{2\,}\left(\frac{c_{\phi}(n)+c_{m^{2}}(n)}{2c_{\phi}(n)+c_{g}(n)}\right)\\
\gamma^{\star}(n) & =\frac{1}{2\,}\frac{g_{1}^{\star}(n)}{(4\pi)^{\frac{n}{2}}}c_{\phi}(n)=\frac{\epsilon}{2\left(2+\frac{c_{g}(n)}{c_{\phi}(n)}\right)}
\end{align*}
this leads to,
\[
\gamma_{m}^{\star}(n)=\left\{ \begin{array}{cc}
\frac{\text{\ensuremath{\epsilon}}}{4} & n=2\\
\frac{\text{\ensuremath{\epsilon}}}{4} & n=3\\
\frac{\epsilon}{6} & n=4\\
\frac{1}{32}\epsilon\left(\frac{5M^{2}}{\text{m0}^{2}}+8\right) & n=5
\end{array}\right.\;\;,\gamma^{\star}(n)=\left\{ \begin{array}{cc}
\frac{\text{\ensuremath{\epsilon}}}{8} & n=2\\
\frac{\text{\ensuremath{\epsilon}}}{16} & n=3\\
0 & n=4\\
-\frac{\epsilon}{16} & n=5
\end{array}\right.
\]
\[
\nu(n)=\left\{ \begin{array}{cc}
\frac{1}{2}+\frac{\epsilon}{8}+O\left(\epsilon^{2}\right)\; & n=2\\
\frac{1}{2}+\frac{\epsilon}{8}+O\left(\epsilon^{2}\right)\; & n=3\\
\frac{1}{2}+\frac{\epsilon}{12}+O\left(\epsilon^{2}\right)\; & n=4\\
\frac{1}{2}+\epsilon\left(\frac{5M^{2}}{64\text{m0}^{2}}+\frac{1}{8}\right)+O\left(\epsilon^{2}\right) & n=5
\end{array}\right.\;\;,\eta(n)=\left\{ \begin{array}{cc}
\frac{\text{\ensuremath{\epsilon}}}{4} & n=2\\
\frac{\text{\ensuremath{\epsilon}}}{8} & n=3\\
0 & n=4\\
-\frac{\epsilon}{8} & n=5
\end{array}\right.
\]
It is noted that for $n=5$, the critical exponent $\eta$ is negative.
This violates the unitarity bound $\eta>0$. This bound is obtained
assuming the validity of the conformal algebra relations in $n$-dimensions\footnote{See for example ref.\cite{Qualls:2015qjb}}.
This violation could be avoided by taking $\epsilon<0$ for $n=5$,
but that would make the non-trivial fixed point $g_{1}^{\star}$ be
negative, which corresponds to a unstable theory. The same situation
happens for any $n\geq5$, this is a consequence of the sign change
mentioned at the end of subsection 3.1. This can also be traced back
to the appearance of inverse powers of the Laplacian in the interaction
term ($\beta>0$).

\section{Conclusions and Outlook}

Conclusions and further research motivated by this work are summarized
in the series of remarks given below, 
\begin{itemize}
\item The critical behavior of a non-local scalar field theory is studied.
This theory has a non-local quartic interaction term which involves
a real power $-\beta$ of the Laplacian. The parameter $\beta$ can
be tuned so as to make that interaction marginal for any dimension.
\item From the Wilson renormalization group point of view, there is no restriction
on the range of interactions. Therefore it makes sense to study the
renormalization of a non-local field theory. In this work, this has
been done using the field theoretic version of this procedure. This
was done for the first corrections to the two and four-point functions.
It was shown that the resulting critical theories are consistent for
dimensions $d<5$.
\item For dimensions $d\geq5$ the theories have a critical index $\eta<0$
which implies a violation of the unitarity bounds.
\item It would be interesting to identify concrete statistical models whose
criticality is described by these fixed points and critical exponents.
\end{itemize}
Summarizing, it is believed that the study of non-local field theories
can enlarge our knowledge about the fixed points, and renormalization
group flows in the space of all possible couplings mentioned in the
introduction. In particular, the study presented in this work shows
the existence of new non-trivial fixed points in dimensions $2$ and
$3$.\vspace{0.5cm}

\textbf{Acknowledgements. }

I am deeply indebted to G. Torroba for sharing his expertise on the
renormalization group, and for many enlightening discussions. I also
thank M. Solis Benitez for many valuable discussions on the contents
of this paper.

\bibliographystyle{unsrt}
\addcontentsline{toc}{section}{\refname}\bibliography{../../../adsqcd/nf/Bibliography,../Bibliography}

\end{document}